# Study of alternative ILC final focus optical configurations

Dou Wang[1], Yiwei Wang[1], Philip Bambade[2], Jie Gao[1]
[1]IHEP, Beijing, China
[2]LAL, Paris, France

*Abstract*: The role of the ILC final focus system (FFS) is to demagnify the beam to the sizes at the IP required to meet the ILC luminosity goals. The current design of final focus achieves perfect first order chromaticity correction in both horizontal and vertical planes. Based on this design a set of alternative optical configurations were studied in which the horizontal beam size at the IP is increased while at the same time the vertical one is decreased, with the goal of reducing beamsstrahlung emission. Luminosity reduction due to the hour glass effect must be considered because βy*/σz* becomes much smaller in this case. The beam-beam interaction simulation Guineapig++ was used for evaluation. Reduced bunch lengths are used to make sure that the obtained luminosities are not lower than those of the ILC original (nominal) configuration.

## 1. Introduction

### 1.1 ILC final focus design and local chromaticity correction

The role of the ILC final focus system (FFS) is to demagnify the beam to the sizes at the IP required to meet the ILC luminosity goals [1]. The FFS optics creates a large and almost parallel beam at the entrance to the final doublet (FD) of strong quadrupoles. Since particles of different energies have different focal points, even a relatively small energy spread of ~0.1% significantly dilutes the beam size, unless adequate corrections are applied. The design of the FFS is thus mainly driven by the need to cancel the chromaticity of the FD. The ILC FFS adopts the idea of local chromaticity correction [2] using two sextupoles (SD0, SF1) attached to the final doublets. A bend upstream generates dispersion across the FD, which is required for the sextupoles to cancel the chromaticity. The dispersion at the IP is zero and the angular dispersion is about $\eta_x' \sim 0.009$, i.e. small enough that it does not significantly increase the beam divergence. Half of the total horizontal chromaticity of the whole final focus is generated upstream of the bend in order for the sextupole (SF1) to simultaneously cancel the chromaticity and the second-order dispersion. The sextupoles in FD also generate the second-order geometric aberrations, so two more sextupoles (SD4, SF5) upstream of the bend are required for cancelling the geometric aberrations.

The horizontal and the vertical sextupoles are interleaved in this design, so they generate third-order geometric aberrations. In ILC FFS a fifth sextupole (SF6) which is in proper phase with the FD sextupoles and an additional bend section upstream have been used to decrease the chromaticity through the system and aberrations at the IP. The residual higher order aberrations can in principle be further minimized with octupoles and decapoles, if needed. The ILC final focus optics for nominal design

($\beta x^*/\beta y^*$=15mm/0.4mm) is shown in Fig. 1.

Since the geometry for FD is fixed, any adjustments of the overall magnification must be introduced upstream of the FFS, using six quadrupoles in the so-called matching section (QM16 to QM11). These quadrupoles are also used to allow the matching of the Twiss parameters that comes from the upstream beam lines in the presence of focusing errors.

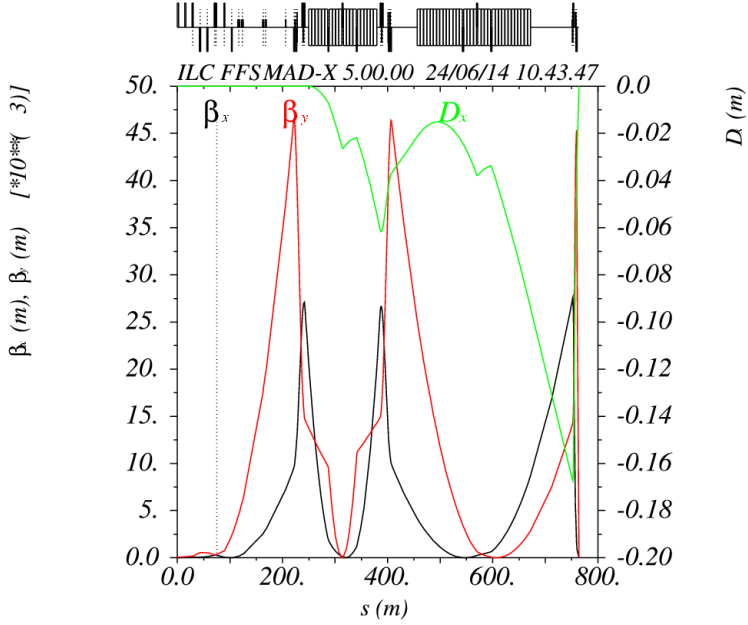

Fig. 1: The ILC final focus optics for nominal design ($\beta x^*/\beta y^*$=15mm/0.4mm).

## 1.2 Luminosity for linear collider

For the linear collider, its luminosity can be estimated analytically according to the parameterization given by Yokoya and Chen [3]:

$$L_0 = \frac{f_{rep} N_b N_e^2}{4\pi \sigma_x^* \sigma_y^*} H_D \tag{1}$$

$$D_{x,y} = \frac{2 N_e r_e}{\gamma} \frac{\sigma_z}{\sigma_{x,y}(\sigma_x + \sigma_y)} \tag{2}$$

$$H_{D_{x,y}} = 1 + D_{x,y}^{1/4} \frac{D_{x,y}^3}{D_{x,y}^3 + 1}[\ln(\sqrt{D_{x,y}}+1) + 2\ln(0.8\frac{\beta_{x,y}}{\sigma_z})] \tag{3}$$

$$H_D = \sqrt{H_{D_x}} H_{D_y}^{(1+2(\frac{\sigma_x}{\sigma_y})^3)/6(\frac{\sigma_x}{\sigma_y})^3} \quad (H_D \approx H_{D_y}^{1/3} \text{ for flat beam}) \tag{4}$$

where $f_{rep}$ is the repetition rate for the bunch train, $N_b$ is the bunch number, $N_e$ is the particle number per bunch, $\sigma_x^*$ and $\sigma_y^*$ are the horizontal and vertical beam size at IP, $H_D$ is the luminosity enhance factor, and $D_{x,y}$ is the disruption factor due to beam-beam pinch effect. We have to point out that these formulae are only valid when the hour glass effect is small enough ($A_y = \frac{\sigma_z}{\beta_y^*} < 1$). Since the validity of these formulae is limited and involve approximation, the resulting luminosity values are only rough estimates. In this note, the beam-beam interaction simulation Guineapig++ [4] is used to give more realistic estimates of the luminosity.

## 2. Alternative design with mainly vertical chromaticity correction

### *2.1 pure vertical chromaticity correction with only two sextupoles*

Firstly, we try to redesign the ILC FFS optics with only two vertical sextupoles SD0 and SD4 while scanning the IP vertical beta function at a large range. The idea of pure vertical chromaticity correction was first developed in reference [5]. This subsection is a crosscheck and continuity of reference [5] with hour glass effect consideration and luminosity simulations. Each time we rematch the IP beta function to a different value using six matching quadrupoles (QM16 to QM11) and then use SD0 and SD4 to correct the second order vertical chromatic item $T_{342}$ and $T_{346}$ by the code of Madx [6]. The energy spread of the beam which was used in our study is 0.0006. As an example to see the effects of pure vertical chromaticity correction, the beam sizes which is calculated by Mapclass [7] for the case of $\beta x^*$=15mm were shown in Figure 2. It can be seen that the $9^{th}$ order horizontal beam size is much larger than the linear one because there is no chromaticity correction in the horizontal direction. Also, the vertical chromaticity will not be perfect when $\beta y^*$ is smaller than 0.4 mm. Meanwhile we can see there is a minimum for $\sigma y^*$ at $\beta y^*$=0.1mm in Fig. 2(b). When we further to enlarge $\beta x^*$ from 15 mm to 100mm and repeat the study as Fig. 2, we always found a minimum for $\sigma y^*$ at $\beta y^*$=0.1mm. In addition, we found both the $9^{th}$ order beam size and the difference of the $9^{th}$ order beam size and linear one in horizontal direction became smaller while enlarging $\beta x^*$ if $\beta x^*$ is smaller than 60 mm, because we have decreased the horizontal chromaticity by enlarging $\beta x^*$ and also $\sigma x^*$. When $\beta x^*$ is increased larger than 90 mm, the horizontal beam size $\sigma x^*$ will increase again while keeping the difference of the $9^{th}$ order beam size and linear one constant. If compare the minimum vertical beam size obtained in Fig. 2(b) with the results showed in Fig. 4(a) of reference [5], we found the minimum vertical beam size here with only two SD sextupoles is significantly smaller than that for the nominal configuration where all sextupoles were used. Now this allows us to identify why the vertical beam size is limited in the nominal design: it is the presence of the 3 SF sextupoles, via higher order coupling terms such as $T_{313}$, $T_{314}$, $U_{3136}$, $U_{3246}$ and so on. Once we turn off the SF sextupoles, even for 15 mm $\beta x^*$, the impact of such coupling terms are reduced.

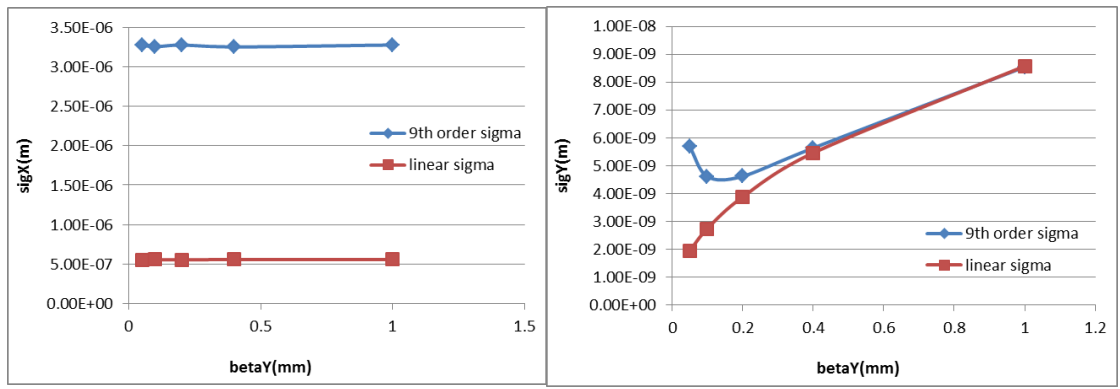

(a) (b)

Fig. 2: (a) The horizontal beam size at IP as a function of βy* when βx*=15mm. (b) The vertical beam size at IP as a function of βy* when βx*=15mm.

For the idea of simplified chromaticity scheme with only two vertical sextupoles, we need to enlarge the horizontal beam size and reduce vertical beam size in order to reduce the horizontal chromaticity and meanwhile keep similar value for σx*×σy* so that to guarantee the geometrical luminosity will not decrease. This can bring in advantages that the beamstrahlung effect has been reduced because the horizontal beam size became larger. Weaker beamstrahlung is good for the physics analysis, which need as narrow as possible a luminosity spectrum, and good to minimize the power losses in the post-IP extraction line. Meanwhile, fewer sextupoles in this scheme also could make the experimental optics tuning easier and faster. On the other hand, the smaller vertical beam size will enhance the hour glass effect and hence decrease the luminosity. So we need to reduce the bunch length to mitigate the hour glass effect and re-check the luminosity at the same time. First, we choose σz=150 μm considering it is the minimum value which ILC may get although it is not easy and need redesign of the bunch compressor system. Then, we chose two critical cases which are βy*=0.1 mm and βy*=0.2 mm respectively. Where βy*=0.1mm is the case for minimum vertical beam size after pure vertical chromaticity correction with SD0 and SD4 and βy*=0.2 aims to keep same hourglass effect as ILC nominal design ($A_y=\sigma_z/\beta_y^*=0.75$). For each fixed βy*, we scanned βx* to find the optimized pair for βx* and βy*, and then did the beam-beam simulations by the code Guineapig++ which is a widely used program on linear collider. Fig. 3(a) shows the value of σx*×σy* with modified IP beta function and pure vertical chromaticity correction. The simulated luminosity by Guineapig++ when σz=150 μm was shown in Fig. 3(b) (The energy spread of both electron beam and positron beam for beam beam simulations is 0.0006.). From Fig. 3 we can get the conclusion that the luminosity is always lower than that of nominal design (40% reduction) with only vertical correction. The reason of luminosity reduction in spite of both the product of σx*×σy* and the ratio of σz/βy* being close to the nominal design is that, according to the formulae from (2) to (4), the shorter bunch length by a half gives much smaller disruption parameter $D_{x,y}$ especially in the vertical plane and hence results in less luminosity enhancement from the pinch effect. Also, it can be seen that there is a rather flat maximum luminosity for the region of 60mm<βx*<90mm because σx* and σy* are almost constant inside this range for βx*.

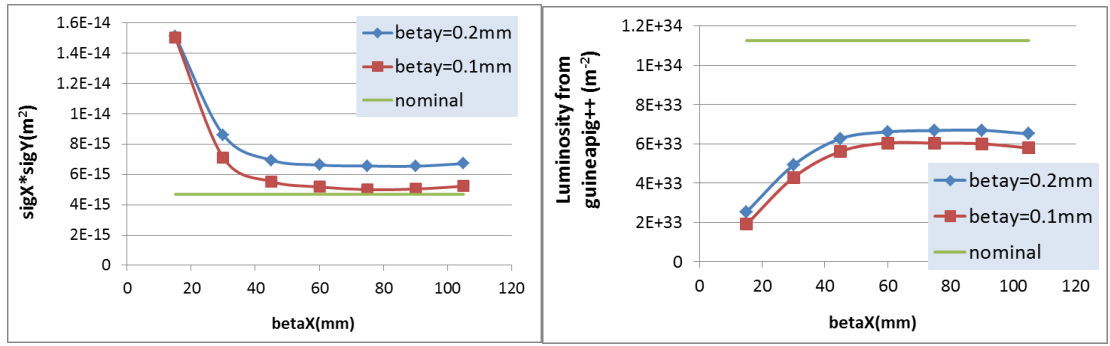

(a) (b)

Fig. 3: (a)The product of IP horizontal beam size and vertical beam size with modified beta function and pure vertical chromaticity correction (green line is the according value for ILC nominal design). (b)The simulated luminosity by Guineapig++ when σz=150 μm (green line is the luminosity for ILC nominal design)[1].

## 2.2 Simplified chromaticity correction scheme with three sextupoles

In order to recover the luminosity drop from the pure vertical correction scheme, we added a horizontal sextupole to the two vertical sextupoles SD0 and SD4 to make the partial correction for the horizontal chromaticity. After the sextupoles refitting with different horizontal sextupoles, we found SF5 is most effective to realize horizontal chromaticity correction and hence to recover the luminosity. As a final summary for the study of simplified chromaticity correction scheme, we proposed two typical alternative designs for ILC FFS in table 1. We have checked the luminosity of the proposal in reference [5] with 75 mm βx* and 0.06 mm βy*. The luminosity of that design is about 39% of nominal design. So with the beam-beam full simulation a more complete optimization was possible and gave somewhat better results.

Table 1: alternative ILC FFS designs with simplified chromaticity correction scheme

|  | ILC nominal | New-1 | New-2 |
|---|---|---|---|
| Sextupoles used | SD0,SF1,SD4,SF5,SF6 | SD0,SD4 | SD0,SD4,SF5 |
| E/beam (GeV) | 250 | 250 | 250 |
| Ne ($\times 10^{10}$) | 2 | 2 | 2 |
| $\sigma_z$ (um) | 300 | 150 | 150 |
| $\beta^*_{x/y}$ (mm) | 15/0.4 | 60/0.2 | 60/0.2 |
| Ay | 0.75 | 0.75 | 0.75 |
| $\sigma^*_{x/y}$ by MAPCLASS (nm) | 594/7.89 | 1689/3.88 | 1524/3.92 |
| $\sigma^*_x \times \sigma^*_y$ (nm$^2$) | 4687 | 6553 | 5974 |
| Luminosity from guineapig++ ($\times 10^{34}$ m$^{-2}$) | 1.126 | 0.668 | 0.741 |

## 3. ILC FFS new optical configurations using 5 sextupoles

In this section, we try to minimize the product of σx*×σy* with fixed βy* and σz (σz=150 μm) using all 5 sextupoles named as SD0, SF1, SD4, SF5 and SF6. Firstly, we chose βy*=0.2mm in order to keep same hourglass effect as ILC nominal design (Ay=σz/βy*=0.75). Fig. 4 and Fig. 5 show the horizontal beam size and vertical beam size at IP after the chromaticity correction with 5 sextupoles. Also, we repeated the study as Fig. 4 and Fig. 5 with slightly different βy*. We found the results for βy*=0.15mm and βy*=0.25mm are almost same as the one for the case of βy*=0.2mm. As a summary, we have plotted the product of σx*×σy* and the simulated luminosity with different value of βy* in Fig. 6. Table 2 listed the strengths of all 5 sextupoles which are needed for chromaticity re-correction. From Fig. 6, it can be seen that one can't get a higher luminosity than ILC nominal design for 150 um bunch length when βx* is larger than 45mm. Anyway it is possible for us to get higher luminosity when βx*<45mm while keeping similar beamstrahlung level as nominal design, or we can get same luminosity as nominal design with much weaker beamstrahlung effect if we just choose 45 mm βx*. However we should notice that the particle distribution will deviate from the Gaussian distribution when βx* is smaller than 45mm due to incomplete chromaticity correction in vertical plane (see Fig. 5) and hence Fig. 6(b) may underestimate the luminosity especially when βx* is smaller than 45mm because we just use RMS beam size to do the simple beam-beam simulations. Finally, we proposed a set of alternative optical parameters for ILC FFS with the weaker beamstrahlung scheme and the higher luminosity scheme in Table 3.

---

[1] The luminosity is just for single collision. Furthermore, we need to multiply the bunch number and the repetition frequency to get the total luminosity. Here and hereafter, we always refer to the single collision luminosity if there is no special explanation.

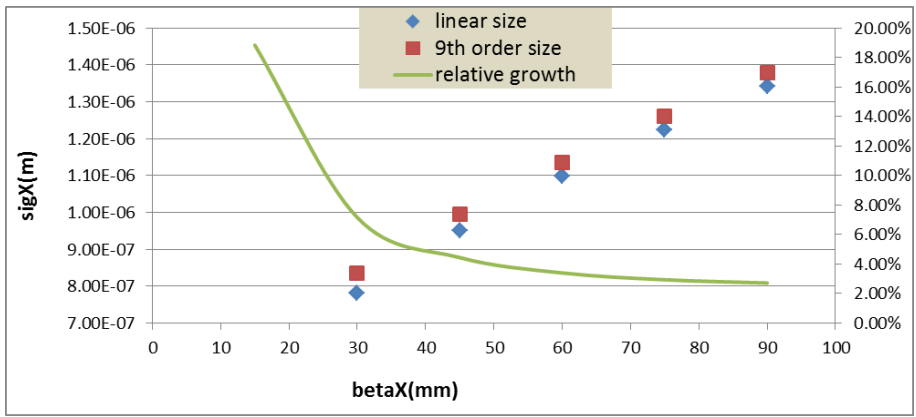

Fig. 4: σx* after the chromaticity correction with 5 sextupoles. $\beta_y^*$=0.2 mm.

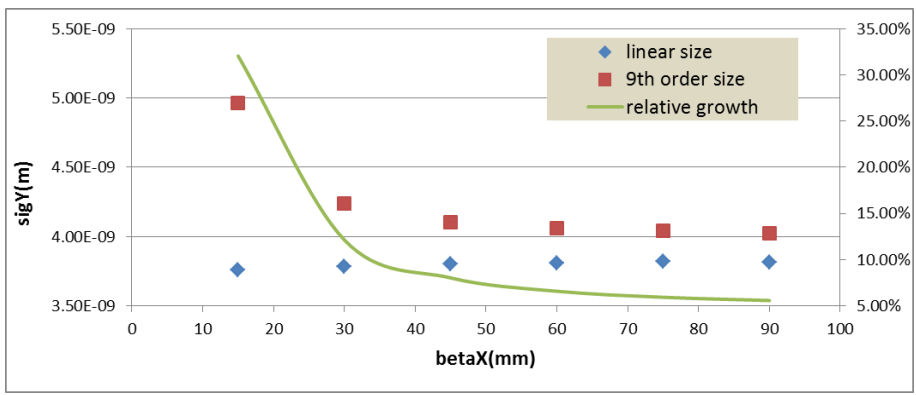

Fig. 5: σy* after the chromaticity correction with 5 sextupoles. $\beta_y^*$=0.2 mm.

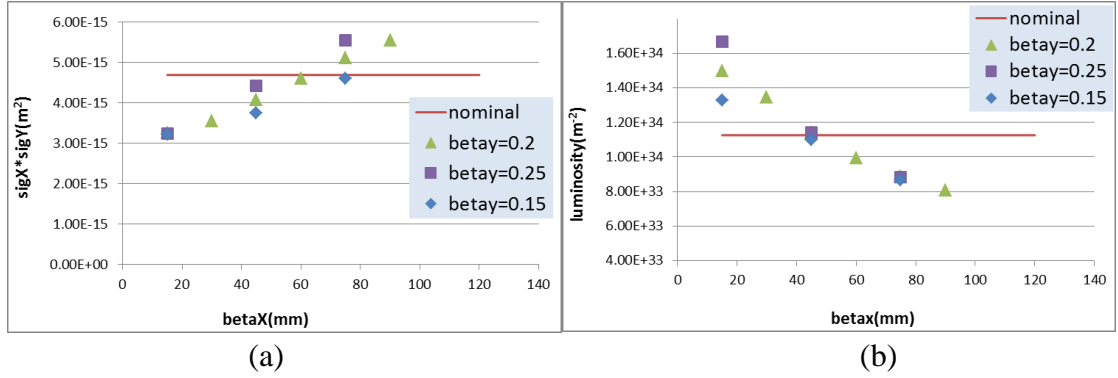

(a) (b)

Fig. 6(a): The product of σx*×σy* as a function of βx* after the chromaticity correction with 5 sextupoles. (b) The luminosity simulated by Guineapig++ as a function of βx* when σz=150 μm.

Table 2: sextupoles' strength needed for chromaticity re-correction

|  | nominal | alternative | | | | | |
|---|---|---|---|---|---|---|---|
|  |  | βy*=0.15mm | | βy*=0.2mm | | βy*=0.25mm | |
| βx/βy [mm] | 15/0.4 | 15/0.15 | 45/0.15 | 15/0.2 | 45/0.2 | 15/0.25 | 45/0.25 |
| SF6 [T/m^2] | 1.668 | -1.326 | -0.748 | -1.273 | -0.806 | -1.365 | -0.845 |
| SF5 [T/m^2] | -0.341 | -2.480 | -1.937 | -2.361 | -1.933 | -2.556 | -2.117 |
| SD4 [T/m^2] | 3.101 | 3.088 | 3.013 | 3.044 | 2.988 | 3.115 | 3.076 |
| SF1 [T/m^2] | -4.959 | -2.092 | -2.376 | -1.924 | -2.212 | -2.182 | -2.545 |
| SD0 [T/m^2] | 7.324 | 7.419 | 7.282 | 7.334 | 7.242 | 7.471 | 7.393 |

Table 3: Alternative optical parameters for ILC FFS with full five sextupoles

|  | ILC nominal | ILC-low BS | ILC-high Lum |
|---|---|---|---|
| E/beam (GeV) | 250 | 250 | 250 |
| Ne (×$10^{10}$) | 2 | 2 | 2 |
| $\sigma_z$ (um) | 300 | 150 | 150 |
| $\beta^*_{x/y}$ (mm) | 15/0.4 | 45/0.2 | 20/0.2 |
| Ay | 0.75 | 0.75 | 0.75 |
| $\sigma^*_{x/y}$ by MAPCLASS (nm) | 594/7.89 | 994/4.10 | 750/4.6 |
| $\sigma^*_x \times \sigma^*_y$ (nm$^2$) | 4687 | 4075 | 3450 |
| Luminosity from guineapig++ (×$10^{34}$ m$^{-2}$) (no waist shift) | 1.126 | 1.143 | 1.40 |
| Beamstrahlung energy spread from guineapig++ (%) | 2.8 | 1.8 | 2.8 |

## 4. Conclusions and future plans

In this note, we have tried both the simplified chromaticity scheme which is mainly in vertical direction with fewer sextupoles (2 or 3) and the thorough chromaticity scheme with 5 sextupoles just as the original FFS design for ILC. Also, the hour glass effect and the final luminosity were studied through beam-beam simulations. With an enlarged σx* and a smaller σy*, we expect less beamstrahlung effect at IP which is an advantage compared with the original design, because it is good for the physics analysis and good to minimize the power losses in the post-IP extraction line.

For 2 sextupoles's (SD0, SD4) correction, we get a simple FFS and much less beamstrahlung, at the expense of much lower luminosity. (The luminosity will drop by about 40% compared with ILC nominal design even with 150 um bunch length.) The luminosity reduction will be 34% with 3 sextupoles (SD0, SD4, SF5). However, if all the 5 sextupoles were used, we can either recover the luminosity with much lower beamstrahlung effect when βx* equals to 45 mm or get a higher luminosity while keeping same beamstrahlung as nominal design with an intermediate horizontal βx* (20mm). Both results are good enough to improve the performance of ILC. However, a smaller bunch length of 150 μm is needed. In the future, we will also try somewhat larger bunch lengths (for example 200 or 250 μm) to see if it would be sufficient to reduce the beamstrahlung while keeping a similar luminosity using the 5 sextupole scheme.